\documentclass[a4paper,pdftex,reqno,11pt]{amsart}

\usepackage{geometry}
\usepackage{graphicx}
\usepackage{courier}
\usepackage{times}
\usepackage{amsmath,amssymb}
\usepackage{hyperref}
\usepackage{color}

\newtheorem{Theorem}{Theorem}[section]
\newtheorem{Proposition}[Theorem]{Proposition}

\newtheorem{Corollary}[Theorem]{Corollary}
\newtheorem{Lemma}[Theorem]{Lemma}

\newenvironment{Proof}
{\begin{trivlist}\item[]{{\sc Proof.}}}{\hfill{$\square$}\noindent\end{trivlist}}

\newtheorem{Definition}[Theorem]{Definition}

\newtheorem{Example}[Theorem]{Example}

\begin{document}

\title{A note on the growth of the dimension in complete simple games}

\author{Sascha Kurz}
\address{\textnormal{Fakult\"at f\"ur Mathematik, Physik und Informatik, Universit\"at Bayreuth, Germany,
email: sascha.kurz@uni-bayreuth.de, tel: +49 921 557353.}
}
\maketitle

{\small \noindent \textbf{Abstract:} The remoteness from a simple game to a weighted game can be measured by the concept of the dimension or the more general Boolean dimension. It is known that both notions can be exponential in the number of voters. For complete simple games it was only recently shown in \cite{o2017growth} that the dimension can also be exponential. Here we show that this is also the case for complete simple games with two types of voters and for the Boolean dimension of general complete simple games, which was posed as an open problem in \cite{o2017growth}. }

{\small \medskip }



\section{Introduction}
\noindent
Simple games can be viewed as binary yes/no voting systems in which a proposal is pitted against the status quo. In the subclass of weighted games each voter has a non-negative weight and a proposal is accepted if the weight sum of its supporters meets or exceeds a preset positive quota. The representation complexity of weighted games is rather low, which makes them interesting candidates for real-world voting system. More precisely, for a weighted game it is sufficient to list the weights of the $n$ voters and the quota. Directly storing whether a proposal would be accepted or rejected for each subset of the $n$ voters would need $2^n$ bits.\footnote{Also the restriction to minimal winning coalitions, which are introduced later on, do not decrease the representation complexity too much, since there can be ${n\choose{\left\lfloor n/2\right\rfloor}}$ minimal winning coalitions.} However, each simple game can be written as the intersection of a finite number of weighted games and the smallest possible number is called the dimension of the simple game. Unfortunately, the dimension can also be exponential in the number of voters, see e.g.\ \cite{korshunov2003monotone,taylor1999simple}. Complete simple games lie in between the classes of simple and weighted games. Here the voters do not admit weights but are completely ordered (which will be defined more precisely in the next section). E.g.\ the voting rules of the Council of the European Union according to the Treaty of Lisbon can be modeled as a non-weighted complete simple game. In \cite{o2017growth} it was shown that the dimension of a complete simple game can also be exponential in the number of voters. However, the stated construction requires that the number of types of different voters also increases without bound. Here we show that the dimension of a complete simple game can be also exponential in the number of voters for just two different types of voters. If all voters are of the same type, then the game is weighted, i.e., has a dimension of 1. The concept of the dimension and the intersection of weighted games was generalized to, more general, Boolean combinations of weighted games, see e.g.\ \cite{faliszewski2009boolean,korshunov2003monotone}. For simple games the corresponding Boolean dimension can also be exponential in the number of voters, see e.g.\ \cite{faliszewski2009boolean,korshunov2003monotone}. Whether the Boolean dimension of a complete simple game can also be exponential in the number of voters was posed as an open problem in \cite{o2017growth}. Here we answer this question by a construction and show that the Boolean dimension is polynomially bounded in the number of shift minimal winning vectors and voters. We also answer another open question from \cite{o2017growth} and analyze possible restrictions on the weights that still allow a representation of a complete simple game as the intersection of weighted games.

The paper is organized as follows. In Section~\ref{sec_preliminaries} we introduce the necessary preliminaries. Our results are presented in Section~\ref{sec_results}.

\section{Preliminaries}
\label{sec_preliminaries}

Let $N=\{1,\dots,n\}$ be a set of $n$ voters. By $2^N$ we denote the set $\{S\,:\, S\subseteq N\}$ of all subsets of $N$. 
We also call the elements $S\in 2^N$ \emph{coalitions}. 

\begin{Definition}
  A \emph{simple} game is a mapping $v\colon 2^N\to\{0,1\}$ such that $v(\emptyset)=0$, $v(N)=1$, and $v(S)\le v(T)$ for all 
	$\emptyset\subseteq S\subseteq T\subseteq N$. Each coalition $S\subseteq N$ with $v(S)=1$ is called a \emph{winning coalition} 
	and each coalition $T\subseteq N$ with $v(T)=0$ is called a \emph{losing coalition}. If $S$ is a winning coalition and all 
	proper subsets of $S$ are losing, then $S$ is called a \emph{minimal winning coalition}. Similarly, we call a losing coalition 
	a \emph{maximal losing coalition} if all proper supersets are winning. Given a simple game $v$, we denote the set of minimal 
  winning coalitions by $\overline{\mathcal{W}}$ and the set of maximal losing coalitions by $\overline{\mathcal{L}}$.
\end{Definition}  

A simple game is uniquely characterized by either its set $\overline{\mathcal{W}}$ of minimal winning coalitions or its set 
$\overline{\mathcal{L}}$ of maximal losing coalitions.

\begin{Example}
  \label{ex_1}
  For $n=4$ voters let $v$ be the simple game with $\overline{\mathcal{W}}=\big\{\{1,2\},\{3,4\}\big\}$. The corresponding set 
  of winning coalitions is given by $\overline{\mathcal{W}}$ and all coalitions of cardinality at 
  least $3$. We have $\overline{\mathcal{L}}=\big\{\{1,3\},\{1,4\},\{2,3\},\{2,4\}\big\}$ and the other losing coalitions are 
  those coalitions of cardinality at most $1$. 
\end{Example}

\begin{Definition}
	Given a simple game $v$, we write $i\sqsupset j$ (or $j \sqsubset i$) for two voters $i,j\in N$ if we have $v\Big(\{i\}\cup S\backslash\{j\}\Big)\ge v (S)$ for all $\{j\}\subseteq S\subseteq N\backslash\{i\}$ and we abbreviate $i\sqsupset j$, $j\sqsupset i$ by $i\square j$. The simple game $v$ is called \emph{complete} (simple game) if the binary relation $\sqsupset$ is a total (complete) preorder, i.e.\
\begin{itemize}
  \item[(1)] $i\sqsupset i$ for all $i\in N$,
  \item[(2)] either $i\sqsupset j$ or $j\sqsupset i$ (including ``$i\sqsupset j$ and $j\sqsupset i$'') for all $i,j\in N$, and
  \item[(3)] $i\sqsupset j$, $j\sqsupset h$ implies $i\sqsupset h$ for all $i,j,h\in N$
\end{itemize}
holds.
\end{Definition}

We remark that the simple game from Example~\ref{ex_1} is not complete.  I.e., while we have $1\square 2$ and $3\square 4$, for each $i\in\{1,2\}$ and each $j\in\{3,4\}$ we have neither $i\sqsupset j$ nor $i \sqsubset j$. 

Since $\square$ is a equivalence relation we can partition the set of voters $N$ into subsets $N_1,\dots,N_t$ such that we have $i\square j$ for all $i,j\in N_h$, where $1\le h\le t$, and $i\square j$
implies the existence of an integer $1\le h\le t$ with $i,j\in N_h$. We call each set $N_h$ an \emph{equivalence class (of voters)} and $t$ the \emph{number of equivalence classes of voters}. We also say that $v$ has $t$ \emph{types of voters}. By $n_i$ we denote the cardinality of $N_i$, where $1\le i\le t$. Given the equivalence classes of voters, we can associate to each coalition $S\subseteq N$ a vector $\widetilde{m}=\left(m_1,\dots,m_t\right)\in\mathbb{N}^t$ via $m_i=\#\left(S\cap N_i\right)$ for all $1\le i\le t$. We also call $\widetilde{m}$ the \emph{type} of coalition $S$. While several coalitions can be associated to the same vector, i.e.\ have the same types, they are either all winning or all losing, so that we speak of winning or losing vectors, respectively.

\begin{Definition}
	Let $v$ be a complete simple game with equivalence classes $N_h$ of voters for $1\le h\le t$. We call a vector $\widetilde{m}=(m_1,\dots,m_t)\in\mathbb{N}^t$, where $0\le m_h\le n_h$ for $1\le h\le t$, a \emph{winning vector} if $v(S)=1$, where $S$ is an arbitrary subset of $N$ containing exactly $m_h$ elements of $N_h$ for $1\le h\le t$. Analogously, we call $\widetilde{m}$ a \emph{losing vector} if $v(S)=0$.
\end{Definition}
W.l.o.g.\ we will always assume that for a complete simple game the equivalence classes of voters are ordered such that we have $l\sqsupset l'$ for all $l\in N_i$ and all $l'\in N_j$ with $1\le i<j\le t$.

The minimal winning vectors for Example~\ref{ex_1} are $(2,0)$ and $(0,2)$. There is a unique maximal losing vector $(1,1)$ and the additional losing vectors are given by $(0,0)$, $(1,0)$, and $(0,1)$. 

\begin{Definition}
	Let $v$ be a complete simple game. We call a minimal winning coalition $S$ \emph{shift minimal} if for every other winning coalition $S'$ with $\{i\}=S\backslash S'$ and $\{j\}=S'\backslash S$ we have $i\sqsubset j$. Similarly, we call a maximal losing coalition $T$ \emph{shift maximal} if for every other losing coalition $T'$ with $\{i\}=T\backslash T'$ and $\{j\}=T'\backslash T$ we have $i\sqsupset j$. Now let $S$ be an arbitrary coalition and $\widetilde{m}=\left(\#\left(S\cap N_1\right),\dots,\#\left(S\cap N_t\right)\right)$ be the corresponding vector. We call $\widetilde{m}$ \emph{shift minimal winning} if $S$ is shift minimal winning and we call $\widetilde{m}$ \emph{shift maximal losing} if $S$ is shift maximal losing. 
\end{Definition}

In words, a coalition is a shift minimal winning coalition, if the coalition is minimal winning and the replacement of any voter by a strictly {\lq\lq}weaker{\rq\rq} (according to $\sqsubset$) voter turns the coalition into a losing one.

Let $v$ be a complete simple game with $t$ equivalence classes of voters. Based on the assumed ordering of the equivalence classes $N_1,\dots,N_t$ we write 
$\widetilde{a}:=\left(a_1,\dots,a_t\right)\succeq \left(b_1,\dots,b_t\right)=:\widetilde{b}$ if $\sum_{h=1}^i a_h\ge \sum_{h=1}^i b_h$ for all $1\le i\le t$. Assume $\widetilde{a}\succeq\widetilde{b}$. If $\widetilde{b}$ is a winning vector, then also $\widetilde{a}$ has to be a winning vector, while it can happen that $\widetilde{b}$ is losing and $\widetilde{a}$ is winning. So, if $\widetilde{m}$ is a winning vector in $v$, then there exists a shift minimal winning vector $\widetilde{m}'$ in $v$ such that $\widetilde{m}\succeq \widetilde{m}'$.

As an abbreviation, we write $\widetilde{a}\succ \widetilde{b}$ if $\widetilde{a}\succeq \widetilde{b}$ and $\widetilde{a}\neq \widetilde{b}$. Note that we can have $\widetilde{a}\succeq\widetilde{b}$ and $\widetilde{b}\succeq\widetilde{a}$ if and only if $\widetilde{a}=\widetilde{b}$.

\begin{Example}
  \label{ex_2}
	Let $v$ be a simple game with $t=2$ equivalence classes of voters $N_1=\{1,2\}$ and $N_2=\{3,4,5,6\}$ such that a coalition $S$ is winning if $\#\left(S\cap N_1\right)\ge 2$ or $\#S\ge 4$. The minimal vectors of $v$ are given by $(2,0)$, $(1,3)$, and $(0,4)$. Since $i\sqsupset j$ for all $i\in N_1$ and all $j\in N_2$ the simple game $v$ is complete and the shift minimal winning vectors are $(2,0)$ and $(0,4)$. Note that $(1,3)\succ (0,4)$, while we have neither $(2,0)\succeq (1,3)$ nor $(1,3)\succeq (2,0)$ . The unique shift maximal losing vector is given by  $(1,2)$.
\end{Example}

\begin{Definition}
	A simple game $v$ is \emph{weighted} if there exists a \emph{quota} $q\in\mathbb{R}_{>0}$ and \emph{weights} $w_i\in\mathbb{R}_{\ge 0}$, where $1\le i\le n$, such that $v(S)=1$ iff $w(S):=\sum_{i\in S}w_i\ge q$ for every coalition $S\subseteq N$. We also write $v=\left[q;w_1,\dots,w_n\right]$.
\end{Definition}

Since $w_i\ge w_j$ implies $i\sqsupset j$, every weighted game is complete, so that the simple game from Example~\ref{ex_1} is not weighted.

\begin{Definition}
  A sequence of coalitions
  $$
    \mathcal{T}=\left(X_1,\dots,X_j;Y_1,\dots,Y_j\right)
  $$
	of a simple game $v$ is called a \emph{trading transform} of length $j$ if
  $$
    \#\left\{i\,:h\in X_i\right\}=\#\left\{i\,:\,h\in Y_i\right\}
  $$
	for all $h\in N$. A trading transform $\mathcal{T}$ is called a \emph{certificate of non-weightedness} for $v$ if $X_1,\dots,X_j$ are winning and $Y_1,\dots,Y_j$ are losing coalitions.
\end{Definition}
The absence of a certificate of non-weightedness of any length is a necessary and sufficient condition for the weightedness of a simple game $v$, see e.g.\ \cite{taylor1999simple}.

A certificate of non-weightedness for Example~\ref{ex_2} is given by 
$$
  \left(\{1,2\},\{3,4\};\{1,3\},\{2,4\}\right)
$$
and by
$$
  \left(\{1,2\},\{3,4,5,6\};\{1,3,4\},\{2,5,6\}\right)
$$
for Example~\ref{ex_2}.

\begin{Definition}
  Let $v_1,\dots,v_d$ be $d$ simple games with the same set of voters $N$. Their \emph{intersection} $v=v_1\wedge \dots\wedge v_d$ is defined via $v(S)=\min\left\{v_i(S)\,:\, 1\le i\le d\right\}$ for all $S\subseteq N$. Similarly, their \emph{union} $v=v_1\vee\dots\vee v_d$ is defined via $v(S)=\max\left\{v_i(S)\,:\,1\le i\le d\right\}$ for all $S\subseteq N$.
\end{Definition}

It can be easily checked that the intersection and the union of a list of simple games is a simple game itself. It is well known that each simple game can be written as the intersection as well as the union of a finite list of weighted games, see e.g.\ \cite{taylor1999simple}. 

\begin{Definition}
  Let $v$ be a simple game. The smallest integer $d$ such that $v$ is the intersection of $d$ weighted games is called the \emph{dimension} of $v$. Similarly, the smallest number $d$ of weighted games such that $v$ is the union of $d$ weighted games is called the \emph{codimension} of $v$.
\end{Definition}

The simple game of Example~\ref{ex_1} can be written as
$$[2;1,1,2,0]\wedge[2;1,1,0,2]\quad\text{or}\quad[2;1,1,0,0]\vee[2;0,0,1,1].$$  Since we already know that the game is not weighted, both the dimension and the codimension are equal to $2$. For the simple game from Example~\ref{ex_2} we have the representations 
$$[8;5,3,2,2,2,2]\wedge[8;3,5,2,2,2,2]\quad\text{and}\quad[2;1,1,0,0,0,0]\vee[4;1,1,1,1,1,1],$$ so that, again, both the dimension and the codimension are equal to $2$.

A useful criterion for a lower bound for the dimension of a simple game is:
\begin{Lemma}
  \label{lemma_lower_bound_dimension} 
  (\cite[Observation 1]{kurz2016dimension}, \cite[Theorem 1]{o2017growth})\\
  Let $v$ be a simple game and let $T_1,\dots,T_d$ be losing coalitions such that for all $1\le i<j\le j$ there is no weighted game $v^{i,j}$ for which every winning coalition of $v$ is winning in $v^{i,j}$ but $T_i$ and $T_j$ are both losing in $v^{i,j}$. Then, the dimension of $v$ is at least $d$.
\end{Lemma}	

\begin{Definition}
  Let $v_1,\dots,v_d$ be simple games. A \emph{Boolean combination} of $v_1,\dots,v_d$ is given by $v_1\wedge v'$ or $v_1\vee v'$, where $v'$ is a Boolean combination of $v_2,\dots,v_d$. For the special case $d=1$ we say that a simple game is a Boolean combination of itself. The \emph{Boolean dimension} of a simple game $v$ is the smallest integer $d$ such $v$ is a Boolean combination of $d$ weighted games $v_1,\dots, v_d$.
\end{Definition}

In words, the Boolean dimension of a simple game $v$ is the smallest number of weighted games that are needed to express $v$ by a logical formula connecting the weighted games using $\wedge$ and $\vee$. As an example we mention that the voting rules of the Council of the European Union according to the Treaty of Lisbon can be written as $(v_1\wedge v_2)\vee v_3$, where $v_1$, $v_2$, and $v_3$ are suitable weighted games. In \cite{kurz2016dimension} it was shown that the dimension is at least $7$ and the codimension is at least $2000$, so that the Boolean dimension is indeed equal to $3$.

\section{Results}
\label{sec_results}

It is well known that the dimension of a simple game is upper bounded by the number $\#\overline{\mathcal{L}}$ of maximal losing coalitions, see e.g.\ \cite{taylor1999simple}. If there exists an equivalence class with many voters this upper bound can be lowered:

\begin{Lemma}
  \label{lemma_weighted_representation}
	Let $v$ be a simple game with $t$ equivalence classes $N_1,\dots,N_t$ of voters and $1\le i\le t$ be fix but arbitrary. For each maximal losing coalition $S\in\overline{\mathcal{L}}$ let $a(S):=\#\left(S\cap N_i\right)$, $S'=S\backslash N_i$, and the weighted game $v^S=\left[q^S;w^S\right]$ be defined by
\begin{itemize}
  \item $q^S=a+1$;
  \item $w_j=1$ for all $j\in N_i$;
  \item $w_j=a+1$ for all $j\in N\backslash\left(S'\cup N_i\right)$; and 
  \item $w_j=0$ for all $j\in S'$.
\end{itemize}
	With this, the intersection $v'$ of the weighted games $v^S$ for $S\in\overline{\mathcal{L}}$ equals $v$.
\end{Lemma}
\begin{Proof}
	Let $S$ be an arbitrary maximal losing coalition in $v$. Since $w^S(S)=a$ and $q^S=a+1$ we have $v^S(S)=0$, so that $v'(S)=0=v(S)$. Since $v'$ is a simple game any losing coalition of $v$ is also losing in $v'$. Now let $T$ be an arbitrary winning coalition and $S$ be an arbitrary maximal losing coalition in $v$. Since $T\not\subseteq S$ there either exists a voter $j\in T\backslash N_i$ with $j\notin S$ or $\#\left(T\cap N_i\right)\ge a(S)+1$. In both cases we have $w^S(T)\ge a(S)+1=q^S$, so 
	that $v^S(T)=1$. Thus, we have $v'(T)=1=v(T)$, which then implies $v'=v$.	
\end{Proof}
While we have constructed a weighted game $v^S$ for each maximal losing coalition $S$, we have $v^S=v^T$ if $S\backslash N_i=T\backslash N_i$ and $\#\left(S\cap N_i\right)=\#\left(T\cap N_i\right)$.

So, it is indeed possible to represent each complete simple game as the intersection of weighted games where the voters of one arbitrary equivalence class of voters always have equal weights. However, in general it is not possible to restrict the intersection to weighted games respecting the strict ordering of the voters:

\begin{Proposition}
  \label{prop_ne}
  There exists a complete simple game $v$ such that for every representation
  $$
    v=\left[q^1;w^1\right]\wedge\dots\wedge\left[q^d;w^d\right]
  $$
  as the intersection of weighted games there exists an index $1\le h\le j$ and two voters $i,j$ from different equivalence classes of voters with $i\sqsupset j$ and $w_i^h<w_j^h$.
\end{Proposition}
\begin{Proof}
	Let $v$ be the complete simple game with $t=4$ equivalence classes of voters, $n_1=n_2=n_3=n_4=20$, and a unique shift maximal losing vector $(4,4,4,4)$. Choose a losing coalition $T\subseteq N$ with $\#\left(T\cap N_p\right)=4$ for all $1\le p\le 4$ and an index $1\le h\le d$ such that coalition $T$ is also losing in $\left[q^h;w^ h\right]$. By eventually scaling the quota $q^h$ and the weights $w^h$ we assume $w^h(T)\le q^h-1$. We set $a_p=w^h\!\left(T\cap N_p\right)/\#\left(T\cap N_p\right)$ and $x_p=w^h\!\left(N_p\backslash T\right)/\#\left(N_p\backslash T\right)$ for all $1\le p\le 4$, i.e., the average weight of members of $T$ or non-members of $T$ in each equivalence class of voters. Note that $(0,9,0,0)$ is a winning vector and choose a coalition $T\cap N_2\subseteq S^1\subseteq N_2$ with cardinality $9$ and minimum weight. With this, $S^1$ is winning and $q^h\le w^h\!\left(S^1\right)\le 4a_2+5x_2$. Since $w^h(T)=4a_1+4a_2+4a_3+4a_4\le q^h-1$, we have
$$
  4a_2+5x_2\ge 4a_1+4a_2+4a_3+4a_4+1,
$$
which is equivalent to
\begin{equation}
  \label{ie_prop_ne_1}
  5x_2 -4a_1-4a_3-4a_4 \ge 1.
\end{equation}
	Note that $(0,0,0,17)$ is a winning vector and choose $T\cap N_4\subseteq S^2\subseteq N_4$ with cardinality $17$ and minimum weights. With this, $S^2$ is winning and $q^h\le w^h\!\left(S^2\right)\le 4a_4+13x_4$.
Since $w^h(T)=4a_1+4a_2+4a_3+4a_4\le q^h-1$, we have
$$
  4a_4+13x_2\ge 4a_1+4a_2+4a_3+4a_4+1,
$$
which is equivalent to
\begin{equation}
  \label{ie_prop_ne_2}
	\frac{13}{4}x_4 -a_1-a_2-a_3 \ge \frac{1}{4}.
\end{equation}
Assuming that $w^h_i\ge w^h_j$ for all voters $i,j$ from different equivalence classes with $i\sqsupset j$, we especially have $a_1\ge x_2$ and $a_3\ge x_4$, which is equivalent to
\begin{equation}
  \label{ie_prop_ne_3}
  5a_1-5x_2 \ge 0
\end{equation}
and 
\begin{equation}
  \label{ie_prop_ne_4}
  5a_3-5x_4\ge 0.
\end{equation}	
	Adding the left and the right hand sides of inequalities~(\ref{ie_prop_ne_1})-(\ref{ie_prop_ne_4}) yields
$$
-\frac{7}{4}x_4 -a_2-4a_4 \ge 1.25,
$$
	which is a contradiction, since $a_2,a_4,x_4\ge 0$. Thus, there exist voters $i$ and $j$ from different equivalence classes of voters with $w_i^h<w_j^h$ and $i\sqsupset j$. 
\end{Proof}

Proposition~\ref{prop_ne} gives a negative answer to the the second question from the conclusion of \cite{o2017growth}, where it is additionally assumed that in an arbitrary equivalence class of voters all weights are equal. The first question from the conclusion of \cite{o2017growth} concerns the worst case behavior of the Boolean dimension of a complete simple game.

A lower bound for worst-case Boolean dimension of a simple or a complete simple game with $n$ voters can be concluded from a simple counting argument. First note that there are at least 
\begin{equation}
  \label{num_simple_games_lower}
  2^{{n\choose {\left\lfloor n/2\right\rfloor}}}> 
	2^{\frac{1}{\sqrt{2\pi n}} 2^n} 
\end{equation}
simple games with $n$ voters, see e.g.\ \cite{korshunov2003monotone} for tighter estimates, and at least
\begin{equation}
  \label{num_complete_simple_games_lower}
	2^{\left(\sqrt{\frac{2}{3}\pi}\cdot 2^n\right)/\left(n\sqrt{n}\right)}
\end{equation}
complete simple games with $n$ voters, see \cite{peled1985polynomial}. 

However, there are not too many possibilities for Boolean combinations:
\begin{Proposition} (\cite[Proposition 1]{faliszewski2009boolean})\\
  The total number of Boolean combinations of $s$ weighted games with $n$ voters is at most $2^{O\left(sn^2\log(sn)\right)}$.
\end{Proposition}
So, as observed in \cite[Corollary 2]{faliszewski2009boolean} and \cite{korshunov2003monotone}, the Boolean dimension of a simple game with $n$ voters can be exponential in $n$. Actually, almost all simple games have an exponential dimension. Using the same reasoning we can also conclude that the Boolean dimension of a complete simple game can be exponential in the number of voters. This answers an open question from \cite{o2017growth}, where it was shown that the dimension of a complete simple game can be exponential in the number of voters.

\begin{Lemma}
  \label{lemma_upper_bound_boolean_dimension_csg}	
  Let $v$ be a complete simple game with $t$ equivalence classes $N_1,\dots,N_t$ of voters. 
	If $v$ has exactly $r$ shift minimal winning vectors $\widetilde{m}^1,\dots,\widetilde{m}^r\in\mathbb{N}^t$, then the Boolean dimension of $v$ is at most $rt$.
\end{Lemma}
\begin{Proof}
	For each index $1\le i\le r$ we will define a complete simple game $v^i$ as the intersection of $t$ weighted games whose unique minimal winning vector coincides with $\widetilde{m}^i$. The union of those $v^i$ give a representation of $v$ as a Boolean combination of $rt$ weighted games.

So, let $\widetilde{m}^i=\left(m_1^i,\dots,m_t^i\right)$. With this, we define the weighted games $v^{i,j}=\left[q^{i,j},w^{i,j}\right]$ by 
\begin{itemize}
  \item $q^{i,j}=\sum\limits_{h=1}^j m_h^i$; 
  \item $w_l^{i,j}=1$ if $l\in \cup_{h=1}^j N_h$ and $w_l=0$ otherwise. 
\end{itemize}
	Now let $v^i=v^{i,1}\wedge\dots\wedge v^{i,t}$. First we check $l\sqsupset l'$ for all $l\in N_h$ and all $l'\in N_{h'}$ with $1\le h<h'\le t$, i.e., the simple game $v^i$ is complete. If $\widetilde{m}=\left(m_1,\dots,m_t\right)\in\mathbb{N}^t$ is a winning vector in $v^i$, then we have $\widetilde{m}\succeq \widetilde{m}^i$. So, if $\widetilde{m}=\left(m_1,\dots,m_t\right)$ is a winning vector in $v'$, then there exists an index $1\le i\le r$ such that $\widetilde{m}\succeq\widetilde{m}^i$. Since that is exactly the condition for $\widetilde{m}$ being a winning vector in $v$, we have $v'=v$. 
\end{Proof}

\begin{Corollary}
  The Boolean dimension of a complete simple game $v$ with $n$ voters and $t$ equivalence classes of voters is at most $tn^t$. If $t=2$, then the Boolean dimension is at most $\left\lfloor\frac{2}{3}(n+3)\right\rfloor$. 
\end{Corollary}
\begin{Proof}
	The number of possible winning vectors $\left(m_1,\dots,m_t\right)$ is at most $n^t$ since $1\le m_1\le n$ if $t=1$ and $0\le m_i\le n-1$ for all $1\le i\le t$ if $t\ge 2$. Thus, the number $r$ of shift minimal winning vectors also is at most $n^t$, so that we can apply Lemma~\ref{lemma_upper_bound_boolean_dimension_csg} to conclude that the Boolean dimension of $v$ is at most $rt\le tn^t$. For the special case $t=2$ we can conclude $n\ge 3r-3$ from \cite[Lemma 1]{kurz2013dedekind}. Thus, the Boolean dimension of $v$ is at most $\left\lfloor\frac{2}{3}(n+3)\right\rfloor$, again using Lemma~\ref{lemma_upper_bound_boolean_dimension_csg}.
\end{Proof}	
We remark that it is well known that complete simple games with a unique equivalence class of voters, i.e., $t=1$, are weighted. The maximum number $r$ of shift minimal winning vectors of a complete simple game with $n$ voters can indeed be exponential in $n$, see \cite{krohn1995directed} for an exact formula for the maximum value of $r$ (depending on $n$).

Next we want to consider the dimension of complete simple games. We remark that the exact dimension is only known for very few simple games.  
In \cite{olsen2016construction} a large family of simple games was constructed, where the dimension could be determined exactly. This yields an explicit description of a sequence of simple games with dimension $2^{n-o(n)}$.

\begin{Proposition}
  \label{prop_parametric_csg_example}
	Let $d\ge 2$ be an integer and $v^d$ be the complete simple game with $t=2$ equivalence classes of voters, where $n_1=d$ and $n_2\ge 2d$, and $r=2$ shift minimal winning vectors $\widetilde{m}^1=(2,0)$, $\widetilde{m}^2=(0,4)$. Then, the dimension of $v$ is exactly $d$.
\end{Proposition}
\begin{Proof}
	W.l.o.g.\ we number the voters so that $N_1=\{1,\dots, d\}$ and $N_2=\{d+1,\dots,n\}$, where $n=n_1+n_2\ge 3d$. For each $1\le i\le i$ we define a weighted game $v^i=\left[q^i;w^i\right]$ by $q^i=8$, $w^i_i=3$, $w^i_j=5$ for 
	all $j\in N_1\backslash\{i\}$, and $w_j^i=2$ for all $j\in N_2$. Let $S$ be an arbitrary winning coalition of $v$. If $\#\left(S\cap N_1\right)\ge 2$, then $w^i(S)\ge 3+5=8=q^i$ for all $1\le i\le d$. If $\#\left(S\cap N_1\right)\le 1$, then $\#S\ge 4$, so that $w^i(S)\ge \# S\cdot 2\ge 8=q^i$ for all $1\le i\le d$. Thus, every winning coalition of $v$ is also winning in $v^i$ for all $1\le i\le d$. Now let $T$ be a losing coalition of $v$. If $T\cap N_1=\emptyset$, then $\#T\le 3$, so that $w^i(T)=2\cdot \#T\le 6<8=q^i$ for all $1\le i\le d$. If $T\cap N_1\neq\emptyset$, then $T\cap N_1=\{i\}$ for a voter $1\le i\le d$ and $\#\left(T\backslash\{i\}\right) \le 2$, so that $w^i(T)=3+2(\#T-1)\le 7<8=q^i$. Thus, we have $v=v^1\wedge\dots\wedge v^d$, so that the dimension of $v$ is at most $d$.

For the other direction we set $T_i=\{i,d+2i-1,d+2i\}$ for all $1\le i\le d$. Since $T_i\cap N_1=\{i\}$ and $T_i\cap N_2=\{d+2i-1,d+2i\}$ the coalition $T_i$ is losing in $v$, where $1\le i\le d$. For all $1\le i<i'\le i'$
$$
  \mathcal{T}^{i,i'}=\left(\{i,i'\},\{2i-1,2i,2i'-1,2i'\};T_i,T_{i'}\right)
$$
	is a certificate of non-weightedness. Thus, we can apply Lemma~\ref{lemma_lower_bound_dimension} to conclude that the dimension of $v$ is at least $d$.
\end{Proof}

The complete simple games in the prof of Lemma~\ref{prop_parametric_csg_example} generalize the complete simple game from Example~\ref{ex_2}. The argument for the lower bound for the dimension of $v$ is the same as in \cite[Proposition 2]{o2017growth}.

Our next aim is to prove that there exist complete simple games with two equivalence classes of voters whose dimension is exponential in the number of voters. To this end, we have to introduce some notation from coding theory. A (binary) code is a subset $C$ of $\mathbb{F}_2^n$ whose elements $c\in C$ are called \emph{codewords}. The \emph{Hamming weight $\operatorname{wt}(c)$} of a codeword $c\in C$ is the number $\#\left\{1\le i\le n\,:\, c_i\neq 0\right\}$ of non-zero coordinates. The \emph{Hamming distance $d(c,c')$} between two codewords $c,c'\in C$ is the number $\#\left\{1\le i\le n\,:\,c_i\neq c_i'\right\}$ of coordinates where $c$ and $c'$ differ. The \emph{minimum Hamming distance $d(C)$} of a code $C$ is the minimum of $d(c,c')$ for all pairs of different codewords $c,c'\in C$. By $A(n,2\delta;w)$ we denote the maximum cardinality of a code $C$ in $\mathbb{F}_2^n$ with minimum Hamming distance $d(C)\ge 2\delta$ such that all codewords $c\in C$ have Hamming weight $\operatorname{wt}(c)=w$. Those codes are called \emph{constant weight codes}. It is well known that $A(n,2;w)={n\choose w}$ and ${n\choose w}\cdot\tfrac{1}{n}\le A(n,4;w)\le{n\choose {w-1}}\cdot \tfrac{1}{w}$, see e.g.\ \cite{graham1980lower}.

\begin{Theorem}
  \label{thm_t_2_dimension_exponential}
  If $n\ge 4$ is divisible $4$, then there exists a complete simple game $v$ with $n$ voters, $t=2$ equivalence classes of voters, and dimension at least
$$
	A(n/2,4;n/4)\ge {{n/2} \choose {n/4}} \cdot\frac{2}{n} \ge \frac{4\cdot 2^{n/2}}{n^2}.
$$	  
\end{Theorem}
\begin{Proof}
	Let $v$ be a complete simple game with $n=4k$ voters, $t=2$ equivalence classes of voters, $n_1=n_2=2k$, $N_1=\{1,\dots, 2k\}$, $N_2=\{2k+1,\dots,4k\}$, and $r=2$ shift-minimal winning vectors $\widetilde{m}^1=(k,0)$, $\widetilde{m}^2=(0,2k)$.

	Let $C_2$ be a code in $\mathbb{F}_2^{2k}$ with $d:=A(2k,4;k)$ codewords of constant weight $k$ and minimum Hamming distance $d(C_2)\ge 4$. Since $\#C_2\le {{2k}\choose{k-1}}\cdot\tfrac{1}{k}$ we can choose a code $C_1$ in $\mathbb{F}_2^{2k}$ with $A(2k,4;k)\le{{2k}\choose{k-1}}=A(2k,2;k-1)$ codewords of constant weight $k-1$ and minimum Hamming distance $d(C_1)\ge 2$. To each codeword $c\in C_1$ we associate the set $\left\{i\,:\, c_i=1, 1\le i\le 2k\right\}$. This gives $d$ sets $T_i^1\subseteq N_1$. Similarly, we associate to each codeword $c\in C_2$ the set $\left\{i+2k\,:\, c_i=1,1\le i\le 2k\right\}$. This gives $d$ sets $T_i^2\subseteq N_2$. With this, we set $T_i:=T_i^1\cup T_i^2$, so that $\#\left(T^i\cap N_1\right)=k-1$, $\#\left(T^i\cap N_2\right)=k$, and $T^i$ is a losing coalition of $v$ for all $1\le i\le d$.   

	For each pair $(i,j)$ with $1\le i<j\le d$ we consider the two losing coalitions $T_i=T_i^1\cup T_i^2$ and $T_j=T_j^1\cup T_j^2$. Since the codewords of $C_1$ have Hamming distance at least $2$, there exists a voter $a^{i,j}$ with $a^{i,j}\in T_i^1$ and $a^{i,j}\notin T_j^1$. Since the codewords of $C_2$ have Hamming distance at least $4$, there exist two different voters $b^{i,j}_1$, $b^{i,j}_2$ with $b^{i,j}_1,b^{i,j}_2\in T_j^2$ and $b^{i,j}_1,b^{i,j}_2\notin T_j^1$. With this, 
	$$
	  \mathcal{T}^{i,j}=\left(T_i\backslash\left\{a^{i,j}\right\}\cup\left\{b^{i,j}_1,b^{i,j}_2\right\},T_j\backslash\left\{b^{i,j}_1,b^{i,j}_2\right\}\cup\left\{a^{i,j}\right\};T_i,T_j\right)
$$
	is a certificate of non-weightedness. Thus, we can apply Lemma~\ref{lemma_lower_bound_dimension} to conclude that the dimension of $v$ is at least $d$, where $d=A(2k,4;k)=A(n/2,4;n/4)$.
\end{Proof}

Due to Lemma~\ref{lemma_upper_bound_boolean_dimension_csg} the complete simple game constructed in the proof of Theorem~\ref{thm_t_2_dimension_exponential} has a Boolean dimension of at most $4$. A null voter in a simple game $v$ is a voter $i$ such that $v(S)=v(S\backslash\{i\})$ for all $S\subseteq N$. By adding up to three null voters, the construction of Theorem~\ref{thm_t_2_dimension_exponential} gives a complete simple game with $n$ voters at dimension at least $4\cdot 2^{(n-3)/2}/n^2$ for each $n\ge 4$. We remark that the number of complete simple games with $n$ voters and $t=2$ types of voters is $Fib(n+6)-(n^2-4n+8)$, see e.g.~\cite[Theorem 4]{kurz2013dedekind}, where $Fib(n)$ denotes the $n$th Fibonacci number, and at most $\frac{n^5}{15}+4n^4$ of them are weighted, see~\cite[Theorem 5.2]{freixaskurzmss2014}. Nevertheless, there are much more complete simple games with two types of voters than weighted games with two types of voters we cannot directly use this to lower bound the worst-case behavior of the dimension. Given a representation $v=v^1\wedge\dots\wedge v^d$ of a (complete) simple game $v$ with dimension $d$ as the intersection of $d$ weighted games, the partition of the voters into equivalence classes typically differ widely across the weighted games $v^i$.  


\begin{thebibliography}{10}

\bibitem{faliszewski2009boolean}
P.~Faliszewski, E.~Elkind, and M.~Wooldridge.
\newblock Boolean combinations of weighted voting games.
\newblock In {\em Proceedings of The 8th International Conference on Autonomou
s
  Agents and Multiagent Systems-Volume 1}, pages 185--192. International
  Foundation for Autonomous Agents and Multiagent Systems, 2009.

\bibitem{freixaskurzmss2014}
J.~Freixas and S.~Kurz.
\newblock On minimum integer representations of weighted games.
\newblock {\em Mathematical Social Sciences}, 67:9--22, 2014.

\bibitem{graham1980lower}
R.~Graham and N.~Sloane.
\newblock Lower bounds for constant weight codes.
\newblock {\em IEEE Transactions on Information Theory}, 26(1):37--43, 1980.

\bibitem{korshunov2003monotone}
A.~D. Korshunov.
\newblock Monotone boolean functions.
\newblock {\em Russian Mathematical Surveys}, 58(5):929--1001, 2003.

\bibitem{krohn1995directed}
I.~Krohn and P.~Sudh{\"o}lter.
\newblock Directed and weighted majority games.
\newblock {\em Zeitschrift f{\"u}r Operations Research}, 42(2):189--216, 1995.

\bibitem{kurz2016dimension}
S.~Kurz and S.~Napel.
\newblock Dimension of the {L}isbon voting rules in the {E}{U} council: a
  challenge and new world record.
\newblock {\em Optimization Letters}, 10(6):1245--1256, 2016.

\bibitem{kurz2013dedekind}
S.~Kurz and N.~Tautenhahn.
\newblock On {D}edekind’s problem for complete simple games.
\newblock {\em International Journal of Game Theory}, 42(2):411--437, 2013.

\bibitem{olsen2016construction}
M.~Olsen, S.~Kurz, and X.~Molinero.
\newblock On the construction of high-dimensional simple games.
\newblock In {\em Proceedings of the Twenty-second European Conference on
  Artificial Intelligence}, pages 880--885. IOS Press, 2016.

\bibitem{o2017growth}
L.~O’Dwyer and A.~Slinko.
\newblock Growth of dimension in complete simple games.
\newblock {\em Mathematical Social Sciences}, 90:2--8, 2017.

\bibitem{peled1985polynomial}
U.~N. Peled and B.~Simeone.
\newblock Polynomial-time algorithms for regular set-covering and threshold
  synthesis.
\newblock {\em Discrete Applied Mathematics}, 12(1):57--69, 1985.

\bibitem{taylor1999simple}
A.~D. Taylor, W.~S. Zwicker, and W.~Zwicker.
\newblock {\em Simple games: {D}esirability relations, trading,
  pseudoweightings}.
\newblock Princeton University Press, 1999.

\end{thebibliography}

\end{document}